\documentstyle[12pt,epsbox]{article}  

\newcommand{\be}{\begin{eqnarray}}
\newcommand{\ee}{\end{eqnarray}}
\newcommand{\ben}{\begin{eqnarray*}}
\newcommand{\een}{\end{eqnarray*}}
\newcommand{\nn}{\nonumber}

\newcommand{\bfig}{\begin{figure}[h]\begin{center}}
\newcommand{\bfigt}{\begin{figure}[t]\begin{center}}
\newcommand{\bfigb}{\begin{figure}[b]\begin{center}}
\newcommand{\efig}{\end{center}\end{figure}}

\newcommand{\bt}{\begin{tabular}}
\newcommand{\et}{\end{tabular}}
\newcommand{\bminip}{\begin{minipage}}
\newcommand{\eminip}{\end{minipage}}

\newcommand{\f}{\frac}
\newcommand{\s}{\sqrt}
\newcommand{\p}{\partial}
\renewcommand{\le}{\left}
\newcommand{\ri}{\right}
\newcommand{\hs}{\hspace}

\newcommand{\Wa}{{W^\ast}}

\newcommand{\varphia}{{\varphi^\ast}}

\newcommand{\psib}{\bar{\psi}}

\newcommand{\Wt}{\tilde W}

\title{QuasiSupersymmetric Solitons of Coupled Scalar Fields
        in Two Dimensions}
\author{S. ONIZAWA \thanks{E-address: onizawa@serra.sci.ibaraki.ac.jp} \\ \\
Department of Mathematical Sciences, Faculty of Sciences,\\
Ibaraki University, Bunkyo 2-1-1, Mito, 310-8512 Japan}
\date{March 19, 2003}
\begin{document}
\maketitle
  \begin{abstract}
     
We consider solitonic solutions of coupled scalar systems,
 whose Lagrangian has a potential term
 (quasi-supersymmetric potential) \\
 consisting of the square of derivative of a superpotential.
The most important feature of such a theory is that among soliton masses
 there holds a Ritz-like combination rule (e.g. $M_{12}+M_{23}=M_{13}$),
 instead of the inequality ($M_{12}+M_{23}<M_{13}$)
 which is a stability relation generally seen in N=2 supersymmetric theory.
The promotion from N=1 to N=2 theory is considered.

  \end{abstract}
\vspace*{3cm}\hspace*{10cm}
IU-MSTP/59   \newpage
\section{Introduction}
     
The (1+1)-dimensional solitons are worthwhile problems because the dynamics
 of (3+1)-dimensional domain walls are directly connected to them.
In Ref.\cite{MTY}, attractive intersoliton force is concluded to exist,
 since there is found a mass inequality ($M_{12}+M_{23}<M_{13}$),
 where mass $M_{ij}$ refers to a soliton interpolating $i$th and $j$th
 potential minimum.
It was the N=2 supersymmetric model which came from D=4, N=1 supersymmetric
 Wess-Zumino model through dimensional reduction.
In the same model, Ref.\cite{AT} and \cite{PT} pointed out
 that there is a curve of marginal stability in parameter space
 where mass equality ($M_{12}+M_{23}=M_{13}$) holds,
 and absence of the intersoliton force of leading order
 is checked right on the curve.

Therefore it will be interesting if we have, not as exceptions but as a rule,
 marginally stable bound states of solitons
 so that the Ritz-like mass combination rule ($M_{ij}=|H_i-H_j|$,
 where $\{H_i\}$ is some discrete series) is satisfied.
Indeed in the cases which we will investigate in this paper,
 it will be a common feature that continuously degenerate class
 of bound solitons comes out.

As for exact soliton solutions of relativistic nonlinear field equations,
 efforts have been continuing since the seventies.
Very sophisticated calculations are invented to study their properties
 \cite{R}\cite{ST}\cite{hR}.
Nevertheless, even in two dimensions constructing localized solutions
 of two or more coupled fields are not easy
 because general systematic method is unknown.
The invention of supersymmetric theory brought a breakthrough for this purpose
 \cite{WB}.

We will concentrate on those Lagrangian field theory
 whose potential term is a square of derivative of a certain function, 
 the latter function being called superpotential.
In the following, we call them quasi-supersymmetric as this kind of systems
 can be made supersymmetric by introducing the same number of Dirac fields
 according to popular prescription.

In Ref.\cite{MTY}, it was shown that for the D=2, N=2 supersymmetric theory
 of only one superfield, exact soliton solutions are obtainable
 through Hamilton-Jacobi method, and that their masses satisfy
 a triangular mass inequality.
This model, actually, is known to be a perturbed conformal field theory.
By virtue of the solvability, the elastic S-matrix elements
 were also calculated \cite{FMVW}.
In contrast to the N=2 theory
 which is so strongly constrained and has fairly known properties,
 the dynamics of N=1 solitons clearly needs different approach.
In this paper, we investigate such elaborate cases of D=2, N=1 supersymmetry
 in which two scalar fields couple through quasi-supersymmetric potential
\begin{eqnarray*}
  V(\phi) = \f12\sum^2_{i=1}\le[\f\p{\p\phi_i}H(\phi)\ri]^2
\end{eqnarray*}

In Sec. 2, starting from (1+1)-dimensional Lagrangian
 with N=1 supersymmetry, we recapitulate soliton solution.
In Sec. 3, we investigate similar system with N=2 supersymmetry,
 characterized by a holomorphic superpotential.
In Sec. 4, we turn to coupled theory of two scalar fields
 with N=1 supersymmetry, and derive the mass combination rule.
In Sec. 5, the relation of N=1 and N=2 supersymmetry
 in scalar soliton theory is considered.
\section{Solitons of Single Scalar Field with $N=1$ Supersymmetry}
     
We consider the two-dimensional N=1 supersymmetric case \cite{WO}
 in which only one type of supersymmetric multiplet takes place.
The action is 
\begin{eqnarray}
 {\cal S}=\int d^2x\le[-\f 12(\p_\mu\phi)^2-\f 12(H')^2
+\f 12 i\psib\gamma^\mu\p_\mu\psi-\f 12 H''\psib\psi\ri] ,
\end{eqnarray}
where $\phi$ is a real scalar field and $\psi$ is a two component
 Majorana field.
$H(\phi)$ is a real superpotential,
 $H'$ and $H''$ are its derivative with respect to $\phi$.

Equations of motion for $\phi$ and $\psi$ are respectively
\be
&&\p_\mu\p^\mu\phi-H'(\phi)H''(\phi)-\f 12H'''(\phi)\psib\psi=0 
  \label{phi_eq_1}\\ &&i\gamma^\mu\p_\mu\psi-H''(\phi)\psi=0 .
\ee
Minimum points of the scalar potential $V=\f 12(H')^2$ coincide with
 the stationary points of $H$ which satisfy equation
\be
 H'(\phi)=0 .  \label{H'_1}
\ee
If there are $m$ such points $\phi_{(1)}>\phi_{(2)}>\cdots>\phi_{(m)}$,
 they are each candidates for asymptotic values $\phi(\pm\infty)$.

Whenever $m\geq2$, this model has soliton solutions.
A soliton is the solution of equation of motion which take different values
 in $x\rightarrow-\infty$ and $x\rightarrow+\infty$. 
If these values are
\be
 \phi=\phi_{(i)}\hspace{2cm}&{\rm for}&\hspace{0.5cm}x\rightarrow-\infty\nn \\
 \phi=\phi_{(j)}\hspace{2cm}&{\rm for}&\hspace{0.5cm}x\rightarrow+\infty ,
\label{ijsoliton_1}   \ee
 let us call this ($i,j$)-soliton where $i-j=\pm 1$.

We want to derive the Bogomol'nyi bound of the Hamiltonian.
For the static solutions, the Hamiltonian is
\be
 {\cal H}=\int^\infty_{-\infty}dx\le[\f 12(\p_x\phi)^2+\f 12(H')^2\ri] .
  \label{Hamiltonian_1}
\ee
We can deform this Hamiltonian to
\be
{\cal H}=\int^\infty_{-\infty}dx\le[\f 12\{\p_x\phi\mp H'\}^2\pm\p_x H\ri] ,
\ee
where we can use either upper sign or lower sign.
Since the first term is always positive, the lower bound of the Hamiltonian is
 given by the second term.
After integrating the second term over $x$, we obtain the Bogomol'nyi bound
\be
 {\cal H}\geq\pm T_{(i,j)} \label{massb_1}
\ee
where $T_{(i,j)}$ is the difference of $H$ at $\phi_{(j)}$ and $\phi_{(i)}$
\be
 T_{(i,j)}=H(\phi_{(j)})-H(\phi_{(i)}) . \label{topological_1}
\ee
$|T_{(i,j)}|$ may be called the classical soliton mass $M_{ij}$.
On this bound scalar field $\phi$ is saturated by solutions of equation
\be
 \p_x\phi\mp H'=0 .  \label{solitoneq_1}
\ee
We can confirm that solutions of this equation satisfy equation of motion
 (\ref{phi_eq_1}) with $\psi=0$ by differentiating it by $x$
 and using it twice.
\section{Soliton in $N=2$ Supersymmetry}
     
Now we consider the two-dimensional N=2 supersymmetric case
 \cite{MTY}\cite{AT}.
The action is derived from the four-dimensional Wess-Zumino model
via dimensional reduction.
\begin{eqnarray}
{\cal S}=\int d^2x[-(\p_\mu\varphia)(\p^\mu\varphi)-(W')^\ast(W')\nn\hs{2cm}\\
+i\psib\gamma^\mu\p_\mu\psi-\f12W''\psib^c\psi-\f12(W'')^\ast\psib\psi^c] ,
\end{eqnarray}
where $\varphi$ is a complex scalar field and $\psi$ is a Dirac field.
$W(\varphi)$ is a complex, holomorphic superpotential.
A soliton is the solution of the equation of motion which takes
 different values for $x\rightarrow-\infty$ and for $x\rightarrow+\infty$. 
This is similar to the above section,
 except now $\varphi$ is a complex scalar field.
Therefore minimum points of the scalar potential
 $V=(W')^\ast(W')$ are distributed on the complex plane of $\varphi$.
Solutions of $V=0$ coincide with those of
\be    W'(\varphi)=0 .  \label{H'_f}\ee
Suppose that this equation has $m$ solutions
 $\varphi_{(1)}$ , $\varphi_{(2)}$ , $\cdots$ , $\varphi_{(m)}$.
When $i\neq j$ boundary conditions of ($i,j$)-soliton are
\be
&&\varphi=\varphi_{(i)}\hspace{2cm}{\rm for}\hspace{0.5cm}x\rightarrow-\infty
           \nn \\
&&\varphi=\varphi_{(j)}\hspace{2cm}{\rm for}\hspace{0.5cm}x\rightarrow+\infty .
\label{ijsoliton_f}  \ee

We will derive the Bogomol'nyi mass bound in this model.
Equations of motion for $\varphi$ and $\psi$ are respectively
\be
&&\p_\mu\p^\mu\varphi-W'(W'')^\ast-\f12(W''')^\ast\psib\psi^c=0 
  \label{phi_eq_f}\\ &&i\gamma^\mu\p_\mu\psi-(W'')^\ast\psi^c=0 .
\ee

When we consider the static solution, the Hamiltonian becomes
\be
{\cal H}=\int^\infty_{-\infty}dx[(\p_x\varphia)(\p_x\varphi)+(W')^\ast(W')] .
\ee
With arbitrary phase $\alpha$, we can express this as 
\be
{\cal H}=\int^\infty_{-\infty}dx\le[|\p_x\varphi-e^{i\alpha}(W')^\ast|^2
           +\p_x\{e^{-i\alpha}W+e^{i\alpha}\Wa\}\ri] .
\ee
Since the first term is always positive, a lower bound of the Hamiltonian is
 given by the second term.
For ($i,j$)-soliton we have
\be
{\cal H}\geq 2\Re[e^{-i\alpha}T_{(i,j)}] ,  \label{massb_f}
\ee
 where $T_{(i,j)}$ is a topological charge
\be
T_{(i,j)}=W(\varphi_{(j)})-W(\varphi_{(i)}) .  \label{topological_f}
\ee
For this lower bound, scalar field $\varphi$ is saturated
 by solutions of equation
\be
\p_x\varphi-e^{i\alpha}(W')^\ast=0 . \label{solitoneq_f}
\ee
We can confirm solutions of this equation satisfy equation of motion
 (\ref{phi_eq_f}) with $\psi=0$ by differentiating it by $x$
 and using its complex conjugate.

Multiplying this equation by $W'$ and integrating over $x$, we have
\be
T_{(i,j)}=e^{i\alpha}\int^\infty_{-\infty}dx|W'|^2 .
\ee
Thus $e^{i\alpha}$ equal the phase of $T_{(i,j)}$.
If we take
\be
T_{(i,j)}=W(\varphi_{(j)})-W(\varphi_{(i)})
    =e^{i\theta}|W(\varphi_{(j)})-W(\varphi_{(i)})| , \label{phase_f}
\ee
 the only condition in which solutions of equation (\ref{solitoneq_f})
 become finite is
\be
e^{i\alpha}=e^{i\theta} .
\ee
Then mass bound (\ref{massb_f}) is
\be
{\cal H}\geq 2|W(\varphi_{(j)})-W(\varphi_{(i)})| .
\ee
Thus the mass bound is twice the distance of two points on complex $W$ plane.

Now let us replace
\be
e^{-i\theta}W=\Wt  \label{Wt_f}
\ee
and decompose $\varphi=\f1{\s2}(\phi_R+i\phi_I)$ and $\Wt=\f12(\Wt_R+i\Wt_I)$,
where $\phi_R$ and $\Wt_R$ are real part of $\varphi$ and $\Wt$,
 and $\phi_I$ and $\Wt_I$ are imaginary parts respectively.
Further using Cauchy-Riemann equations
\be
&&\f{\p\Wt_R}{\p\phi_R}=\f{\p\Wt_I}{\p\phi_I} \nn \\
&&\f{\p\Wt_R}{\p\phi_I}=-\f{\p\Wt_I}{\p\phi_R} \label{Cauchy-Riemann}
\ee
for (\ref{solitoneq_f}), we have equations
\be
 &&\p_x\phi_R=\f{\p\Wt_R}{\p\phi_R}   \nn \\
 &&\p_x\phi_I=\f{\p\Wt_R}{\p\phi_I} .  \label{solitonWReq_f}
\ee
They mean that the integral curve of Eq.(\ref{solitoneq_f})
 coincides with a gradient curve of $\Wt_R(\phi_R,\phi_I)$.
Thus to derive the finite solution, point $\varphi_{(i)}$
 must be connected with $\varphi_{(j)}$ by a gradient curve of $\Wt_R$.

On the other hand, since $\Wt_R$ is a harmonic function,
 all stationary points are Saddle points.
We illustrate below that a stationary point
 is connected with other point by a gradient curve of $\Wt_R$,
 but it is not connected in general by any gradient curves of $W_R$.

Let us take the cubic superpotential whose general form is
\be
W(\varphi)=a\varphi^3+b\varphi^2+c\varphi,
\ee
where $a,b,c$ are arbitrary complex constants.
When it has two different stationary points,
 by shifting the origin and rescaling we can always arrange it so that
\be
 W(\varphi)=e^{i\beta}\le\{-\f{\s2}3\varphi^3+\f1{\s2}\varphi\ri\} .
   \label{H3_f}
\ee
Here we have maintained a phase $e^{i\beta}$ for convenience.
\bfigt
\psbox[width=9.8cm]{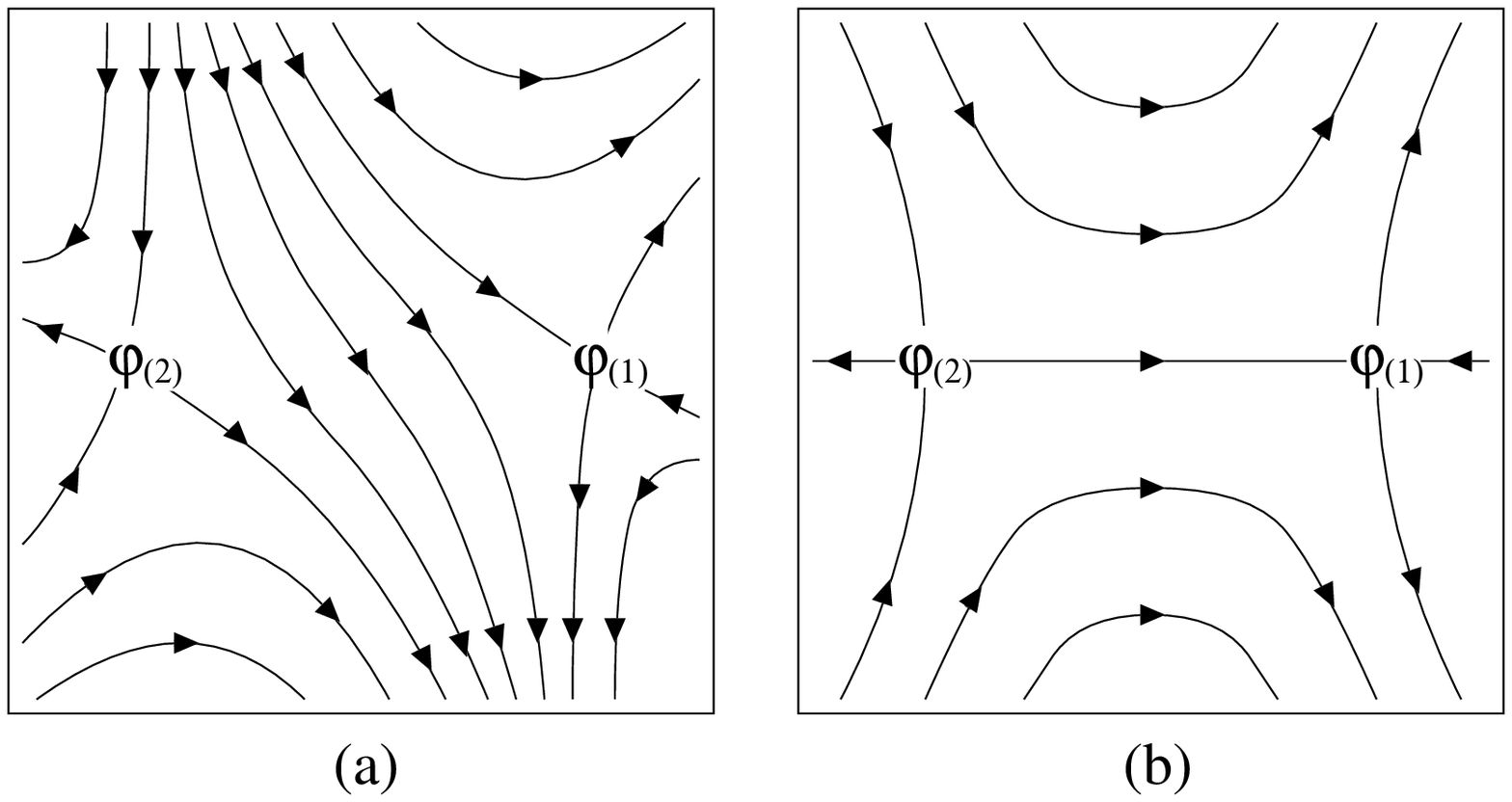}
\caption{(a) Gradient curves of $W_R$ for $\beta=\f\pi4$.
         (b) Gradient curves of $\Wt_R$ for $\theta=\beta$.}\label{fig_Wgrad}
\efig
The real part and the imaginary part of $W$ are
\be
&&W_R=-\f 13\cos\beta{\phi_R}^3+\sin\beta{\phi_R}^2\phi_I
+\cos\beta\phi_R{\phi_I}^2-\f 13\sin\beta{\phi_I}^3 \nn  \\
&&\hs{2cm} +\cos\beta\phi_R-\sin\beta\phi_I  \label{W_R} \\
&&W_I=-\f 13\sin\beta{\phi_R}^3-\cos\beta{\phi_R}^2\phi_I
+\sin\beta\phi_R{\phi_I}^2+\f 13\cos\beta{\phi_I}^3 \nn  \\
&&\hs{2cm} +\sin\beta\phi_R+\cos\beta\phi_I . \label{W_I}
\ee
In this case stationary points of the superpotential are
 $\varphi_{(1)}=+\f1{\s2}$ and $\varphi_{(2)}=-\f1{\s2}$
 while $W(\pm\f1{\s2})=\pm\f13e^{i\beta}$.
Therefore topological charge (\ref{topological_f}) for (2,1)-soliton is
\be
T_{(2,1)}=+\f23e^{i\beta} .
\ee
Comparing this with (\ref{phase_f}) we have
\be
 e^{i\theta}=e^{i\beta} .
\ee 
Thus Eqs.(\ref{solitonWReq_f}) become 
\be
&&\p_x\phi_R=-\phi_R^2+\phi_I^2+1 \nn \\ &&\p_x\phi_I=2\phi_R\phi_I 
\ee
 and (2,1)-soliton solution is
\be
&&\phi_R=\tanh(x-x_0) \nn \\ &&\phi_I=0 .  \label{21soliton_sol_f3}
\ee

Now comparing [Fig.\ref{fig_Wgrad}(a)] and [Fig.\ref{fig_Wgrad}(b)],
 $\varphi_{(1)}$ is not connected with $\varphi_{(2)}$
 by any gradient curves of $W_R$.
On the other hand it is connected by a gradient curve of $\Wt_R$.
This means that Eq.(\ref{Wt_f}) adjusts gradient curves
 to connect these points by adding $W_R$ and $W_I$ with the ratio of $\theta$.
As well the gradient curve which connects these points
 does not depend on $\phi_I$.
 It corresponds to $\phi_I=0$ in soliton solution (\ref{21soliton_sol_f3}).     \label{N2sol}
\section{Mass Combination Rule}
     
This section we consider the case
 where two scalar fields interact with N=1 supersymmetry.
The action, including fermions, is
\begin{eqnarray}
{\cal S}=\int d^2x\sum_{i=1}^2\le[-\f 12(\p_\mu\phi_i)^2-\f 12(H_i)^2
+\f 12 i\psib_i\gamma^\mu\p_\mu\psi_i-\sum_{j=1}^2\f 12 H_{ij}\psib_i\psi_j\ri]
\label{action_2} \end{eqnarray}
 where superpotential $H$ is
 a real function of $\phi_1$ and $\phi_2$.
The subscript $i$ on $H(\phi_1,\phi_2)$ means the derivative
 with respect to $\phi_i$
\be
 H_i(\phi_1,\phi_2)=\f{\p H(\phi_1,\phi_2)}{\p\phi_i}.
\ee
In this case, minimum points of the scalar potential $V=\sum_i\f 12(H_i)^2$
are scattered in the $(\phi_1,\phi_2)$-plane.
These points are solutions of equations
\be
 &&H_1=0 \nn \\  &&H_2=0. \label{H'_2}
\ee
Suppose that there are $m$ solutions
\be
 \phi_{(1)}&=&(\phi_{1(1)},\phi_{2(1)})  \nn \\
 \phi_{(2)}&=&(\phi_{1(2)},\phi_{2(2)})      \\
        &\vdots&                         \nn \\
 \phi_{(m)}&=&(\phi_{1(m)},\phi_{2(m)}). \nn
\ee
Whenever $m\geq2$, equations of motion have soliton solutions.
Boundary conditions of ($i,j$)-soliton are
\be
 (\phi_1,\phi_2)=(\phi_{1(i)},\phi_{2(i)})
       \hspace{2cm}&{\rm for}&\hspace{0.5cm}x\rightarrow-\infty  \nn \\
 (\phi_1,\phi_2)=(\phi_{1(j)},\phi_{2(j)})
       \hspace{2cm}&{\rm for}&\hspace{0.5cm}x\rightarrow+\infty .
\label{ijsoliton_2}  \ee

The Hamiltonian for the static solution is
\be
{\cal H}=\int dx\le[\sum_{i=1}^2\f 12\{\p_x\phi_i\mp H_i\}^2\pm\p_x H\ri].
\label{deform_2}  \ee
We derive the Bogomol'nyi mass bound similar to (\ref{massb_1}),
 where $T_{(i,j)}$ is
\be
T_{(i,j)}=H(\phi_{1(j)},\phi_{2(j)})-H(\phi_{1(i)},\phi_{2(i)}).
\ee
On this bound, ${\cal H}$ is saturated by solutions of equations
\be
 &&\p_x\phi_1=\pm H_1   \nn \\
 &&\p_x\phi_2=\pm H_2.  \label{solitoneq_2}
\ee
They are similar to (\ref{solitonWReq_f}).
Thus again, the integral curve coincides with a gradient curve of $H$.

However, $H$ is not a harmonic function.
Stationary points of $H$ can be classified into three types.
They may be bottom (minimum), top (maximum) or saddle.
A bottom point is the source of gradient curves [Fig.\ref{fig_grad}(a)]
 and a Top point is the sink [Fig.\ref{fig_grad}(b)].
Namely infinite number of gradient curves start from the bottom point,
 and flow into the top point.
\bfigt
 \psbox[width=9.8cm]{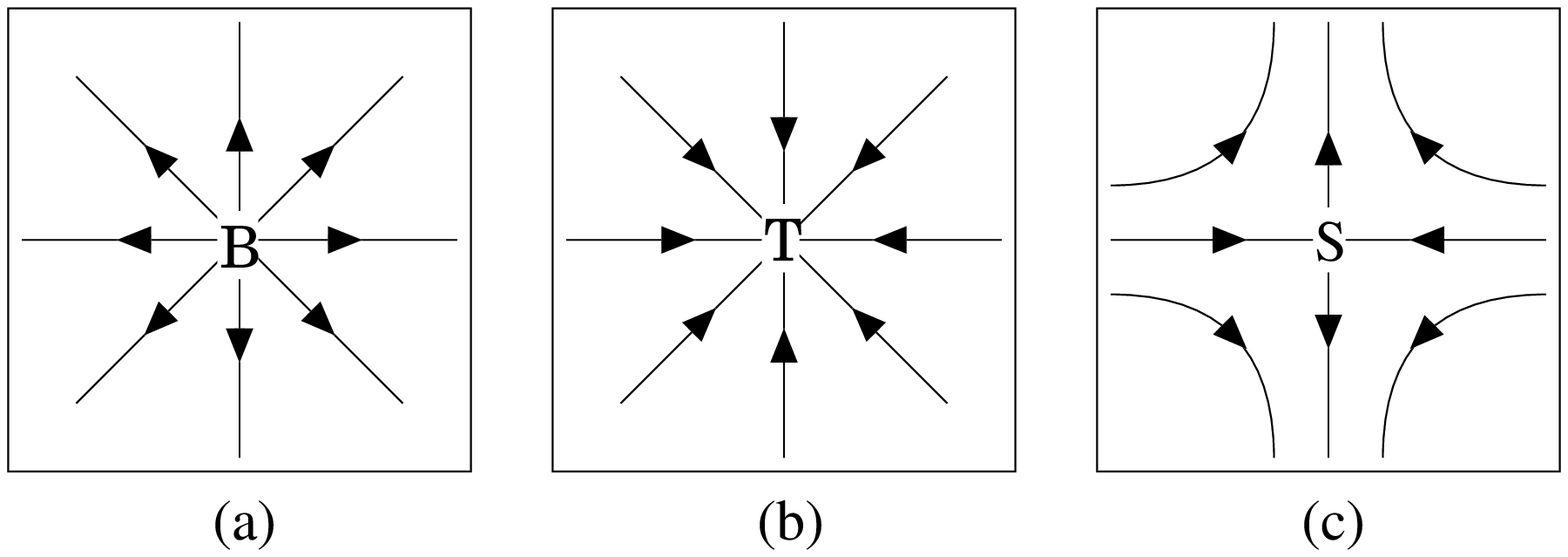}
  \caption{Gradient curves near the (a) bottom, (b) top and (c) saddle point.}
  \label{fig_grad}
\efig
Since the soliton integral curve coincides with a gradient curve,
 if the superpotential has a bottom point and a top point,
 equations of motion may have continuous infinity of soliton solutions.

On the other hand, at the saddle point, only two gradient curves can
 start from it and only two gradient curves come into it
 [Fig.\ref{fig_grad}(c)].
Thus there are at most four soliton solutions which pertain to a saddle point
 on one of the boundary conditions.
In the previous section, since all stationary points of $\Wt_R$
 are saddle, there are only finite number of soliton solutions
 in the N=2 supersymmetry.

Now we consider the cubic superpotential case.
The general form is
\be
H=\sum_{ijk}A_{ijk}\phi_i\phi_j\phi_k+\sum_{ij}B_{ij}\phi_i\phi_j
 +\sum_iC_i\phi_i
\ee
where $A_{ijk},B_{ij},C_i$ are real coefficients,
 and symmetric with respect to the interchange of indices.
Generally in this case, Eqs.(\ref{H'_2}) have four solutions.
Among those four points, we can fix two points at $(\phi_1,\phi_2)=(\pm 1,0)$.
For simplicity we further fix remaining two points
 at $(\phi_1,\phi_2)=(0,\pm p)$, where $p>0$.
Then $H$ has two parameters:
\be
H=-\f 13{\phi_1}^3+a_2{\phi_1}^2\phi_2-a_1\phi_1{\phi_2}^2
+\f 13a_2a_1{\phi_2}^3+\phi_1-a_2\phi_2  \label{H3_2}
\ee
where $a_2$ is an arbitrary constant and $a_1\equiv\f 1{p^2}>0$.
Now let us call $(+1,0)$, $(-1,0)$, $(0,+p)$, $(0,-p)$ as
$\phi_{(1)}$, $\phi_{(2)}$, $\phi_{(3)}$, $\phi_{(4)}$ respectively.

To investigate the aspect near these points,
 we consider the eigenvalues of the Hessian matrix
\be
 H''=\pmatrix{H_{11}&H_{12} \cr  H_{21}&H_{22}} \label{matrix}
\ee
at each stationary point, where
\ben
 &&H_{11}\equiv\f{\p^2H}{\p{\phi_1}^2}=-2\phi_1+2a_2\phi_2    \\
 &&H_{12}\equiv\f{\p^2H}{\p\phi_1\p\phi_2}=2a_2\phi_1-2a_1\phi_2  \\
 &&H_{22}\equiv\f{\p^2H}{\p{\phi_2}^2}=-2a_1\phi_1+2a_2a_1\phi_2 .
\een
Of two eigenvalues, if both are positive, it is a bottom point.
If both are negative, it is a top point.
And if one is positive and the other is negative, it is a saddle point.
See below.
\ben
 \lambda_1<0  ,\ \lambda_2<0 &\cdots & {\rm Top}    \\
 \lambda_1>0  ,\ \lambda_2>0 &\cdots & {\rm Bottom} \\
 \lambda_1\lambda_2<0    &\cdots & {\rm Saddle}
\een
Hereafter we denote the bottom point by B, the top point by T
 and the saddle point by S.

Eigenvalues of matrix (\ref{matrix}) at $\phi_{(1)}$ , $\phi_{(2)}$ ,
 $\phi_{(3)}$ , $\phi_{(4)}$ are respectively
\be
&&\lambda=-(a_1+1)\pm\s{(a_1+1)^2-4(a_1-{a_2}^2)}  \label{eigenvalue1_2} \\
&&\lambda=(a_1+1)\pm\s{(a_1+1)^2-4(a_1-{a_2}^2)} \label{eigenvalue2_2} \\
&&\lambda=a_2(\s{a_1}+\f 1{\s{a_1}})
      \pm\s{{a_2}^2(\s{a_1}+\f 1{\s{a_1}})^2+4(a_1-{a_2}^2)}\\
&&\lambda=-a_2(\s{a_1}+\f 1{\s{a_1}})
   \pm\s{{a_2}^2(\s{a_1}+\f 1{\s{a_1}})^2+4(a_1-{a_2}^2)} .
 \label{eigenvalue4_2}
\ee
The feature of these eigenvalues are dictated by $a_1-{a_2}^2$.
Since $a_1\equiv\f 1{p^2}$ is fixed,
 $a_2$ decides the feature of stationary points of $H$.
The classification of the types of
 ($\phi_{(1)},\phi_{(2)},\phi_{(3)},\phi_{(4)}$) is (Fig.\ref{fig_class})
\be
&&{\rm(S,S,T,B)\hs{1cm} for}\hs{5mm} a_2<-\f 1p       \\
&&{\rm(T,B,S,S)\hs{1cm} for}\hs{5mm} -\f 1p<a_2<\f 1p \\
&&{\rm(S,S,B,T)\hs{1cm} for}\hs{5mm} a_2>\f 1p .
\ee
Values of superpotential $H$ at each point are respectively
 $H(\phi_{(1)})=+\f 23$ , $H(\phi_{(2)})=-\f 23$ ,
 $H(\phi_{(3)})=-\f 23pa_2$ , $H(\phi_{(4)})=+\f 23pa_2$.
We can check $H$ takes the largest value at the top,
 the smallest value at the bottom
 and the intermediate values at the saddle points.
\bfigt
 \psbox[width=8.5cm]{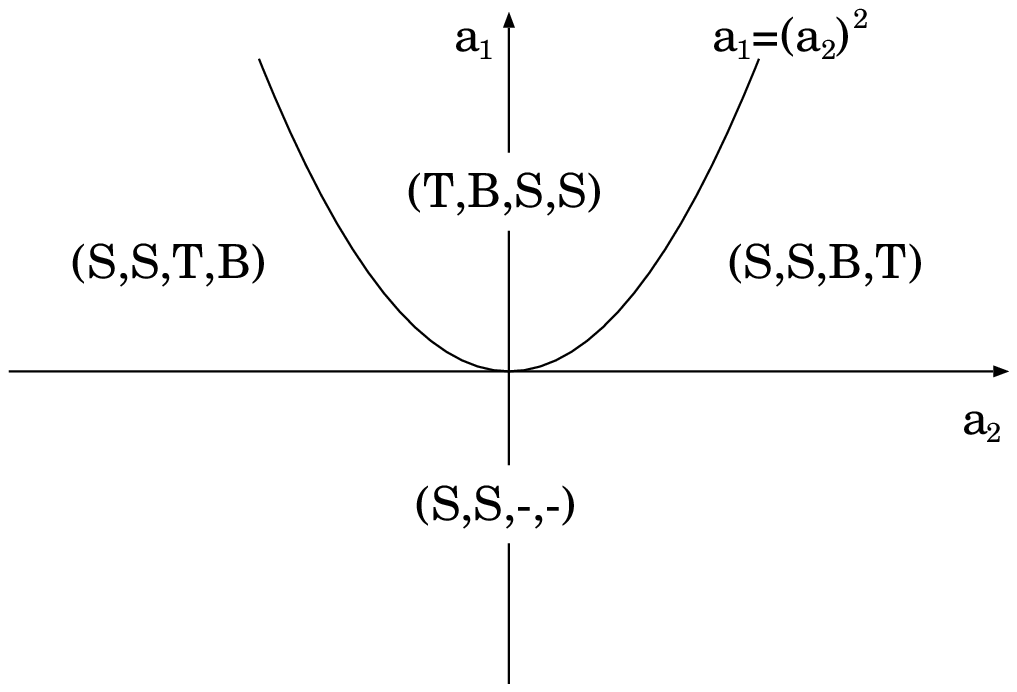}
  \caption{Classification of the type of stability points
            ($\phi_{(1)},\phi_{(2)},\phi_{(3)},\phi_{(4)}$).}
  \label{fig_class}
\efig
\bfigt
 \psbox[width=9.8cm]{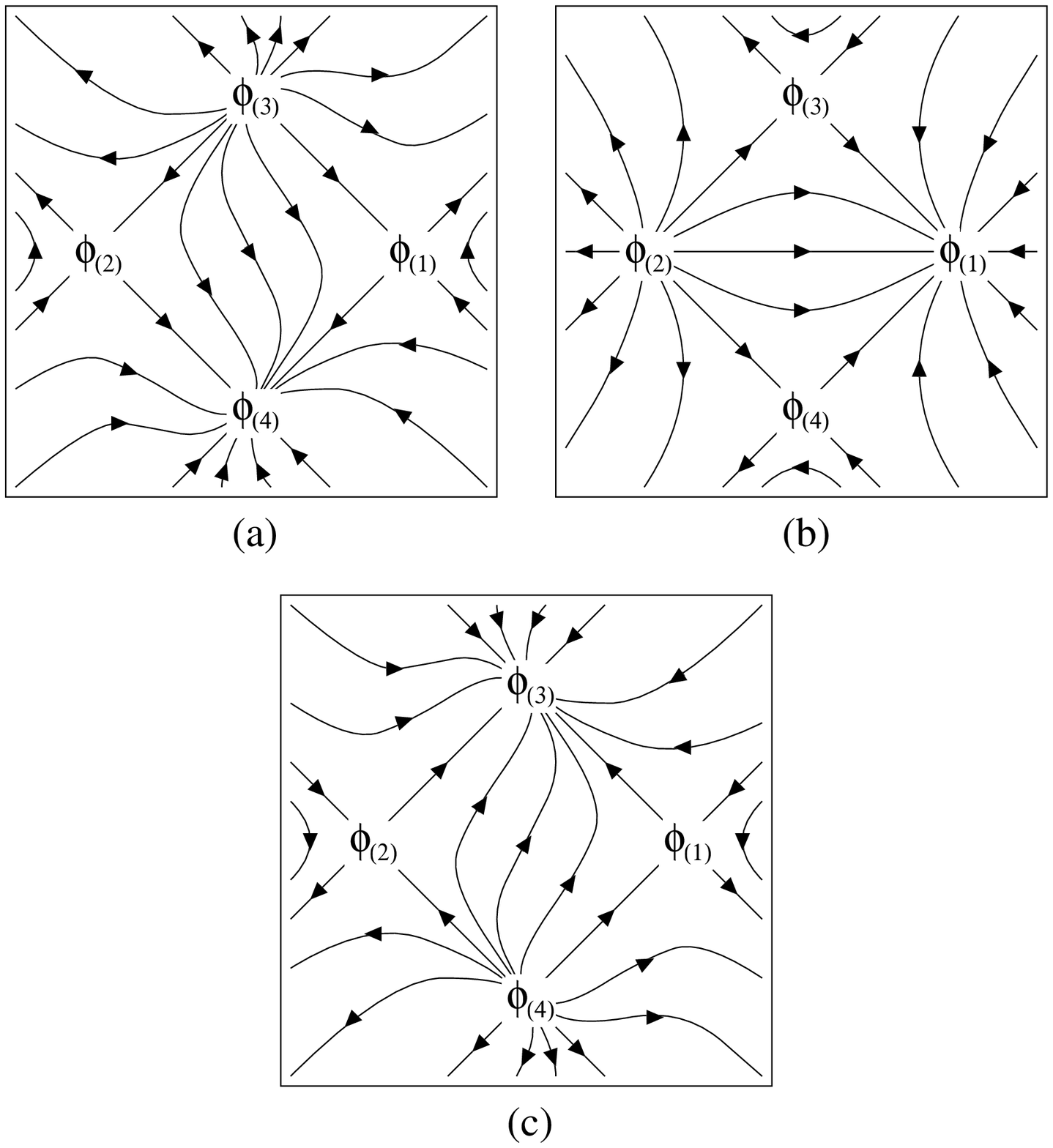}
  \caption{ Gradient curves of $H$ for $a_1=1$.
   (a) (S,S,B,T)-type: $a_2=2$, (b) (T,B,S,S)-type: $a_2=0$,
   (c) (S,S,T,B)-type: $a_2=-2$.}\label{fig_Hgrad}
\efig

Watching (Fig.\ref{fig_Hgrad})
 we find that $H$ have only two gradient curves connecting B and S,
 only two connecting S and T
 and infinite number of gradient curves connecting B and T.
On the other hand it does not have the gradient curve connecting S and S.

As an example, let us examine closely case $-\f1p<a_2<\f1p$
 [Fig.\ref{fig_Hgrad}(b)].
In this case, Eqs.(\ref{solitoneq_2}) have following soliton solutions:
 (4,1)-soliton, (3,1)-soliton, (2,4)-soliton and (2,3)-soliton,
and infinite number of (2,1)-solitons. But there is no (3,4)-soliton.
Further (2,1)-soliton solutions make one parameter family with a common mass.
It may be considered bound states either of 
 (2,3)-soliton and (3,1)-soliton, or (2,4)-soliton and (4,1)-soliton.
And masses of these solitons satisfy an important relation
\be
M_{21}=M_{23}+M_{31}=M_{24}+M_{41} .
\ee
This reminds us of the combination rule of Ritz and implies a relation
 of marginal stability for (1,2)-solitons.
In our special case, masses of these solutions
 are $M_{41}=M_{23}=\f 23(1-pa_2)$ , $M_{31}=M_{24}=\f 23(1+pa_2)$
 and $M_{21}=\f 43$.
In the case of $a_2<-\f1p$ or $a_2>\f1p$, the relation of soliton solutions
 is similar to the case of $-\f 1p<a_2<\f 1p$.

Case $a_1=1$ and ${a_2}^2\neq 1$ is special.
In this case, stationary points of $H$ are
 at $(\phi_1,\phi_2)=(\pm1,0), (0,\pm1)$.
We can solve Eqs.(\ref{solitoneq_2}) as follows.
In terms of new variables
\be
\phi_+=\f 1{\s 2}\{\phi_1+\phi_2\} \\
\phi_-=\f 1{\s 2}\{\phi_1-\phi_2\},
\ee
 superpotential (\ref{H3_2}) becomes
\be
H=\s 2\le\{(1-a_2)\{-\f 13{\phi_+}^3+\f 12\phi_+\}
         +(1+a_2)\{-\f 13{\phi_-}^3+\f 12\phi_-\}\ri\},
\ee
which is the sum of $\phi_+$ part and $\phi_-$ part.
Equation (\ref{solitoneq_2}) reduces to two separate equations
 whose solution is
\be
 \phi_+=\pm\f 1{\s 2}\tanh\le((1-a_2)(x-x_0^+)\ri)  \\
 \phi_-=\pm\f 1{\s 2}\tanh\le((1+a_2)(x-x_0^-)\ri) ,
\ee
where $x_0^+$ or $x_0^-$ is the center of each soliton.
The parameter of the soliton solution curve which connect B and T is
 $\f 12(x_0^+-x_0^-)$. And $\f12(x_0^++x_0^-)$ indicates
 the center of $\phi_+$ soliton and $\phi_-$ soliton.

We have obtained two non-interacting solitons posed at rest.
Each soliton belongs to different degree of freedom: $\phi_+$ and $\phi_-$.
Its obvious generalization to $N$ scalars
 $\phi=(\phi_1, \phi_2, \cdots, \phi_N)$ is
\be
 H(\phi)=\sum^N_{i=1}\le(\f13{\phi_i}^3-\phi_i\ri) .
\ee
Models which we considered in this section
 have many properties in common with this trivial case.
We do not know, however, general criterion
 when this sort of disentanglement of superpotential occurs.
\section{Promotion from N=1 to N=2 Theory}
     
In this section, we will investigate when and how
 the promotion from N=1 to N=2 takes place in two coupled scalar fields.
As the superpotential in N=2 theory is holomorphic,
 we are hinted to force our N=1 superpotential to be harmonic:
\begin{eqnarray}
 \le(\f{\p^2}{\p{\phi_1}^2}+\f{\p^2}{\p{\phi_2}^2}\ri)H=0.
\end{eqnarray}
Then we get $a_1=-1$, and $a_2$ is arbitrary.
In our previous context, this is put that $p$ must be $\pm i$.
It is curious that we obtained harmonic $H$,
 not by pushing $\phi_{(3)}$ and $\phi_{(4)}$ to infinity,
 but by making them imaginary.

Now, in order to understand what happens at $a_1=-1$,
 let us assume $a_1<0$, but not necessary $a_1=-1$.
If $\phi_1$ and $\phi_2$ regard as $\phi_R$ and $\phi_I$ respectively,
 we see that $a_2$ correspond to $\tan\beta$ in (\ref{W_R})
 or $-\f1{\tan\beta}$ in (\ref{W_I}).
Therefore let us replace $a_2$ by $\tan\beta$,
 and rename $H\cos\beta$ as $H$.
\be
&&H=-\f 13\cos\beta{\phi_1}^3+\sin\beta{\phi_1}^2\phi_2
-a_1\cos\beta\phi_1{\phi_2}^2+\f 13a_1\sin\beta{\phi_2}^3 \nn \\
&&\hs{2cm} +\cos\beta\phi_1-\sin\beta\phi_2  \label{H3tanH_2}
\ee
Then all stationary points are saddle points (Fig.\ref{fig_class})
 from (\ref{eigenvalue1_2}) and (\ref{eigenvalue2_2}).
\bfigt
 \psbox[width=9.8cm]{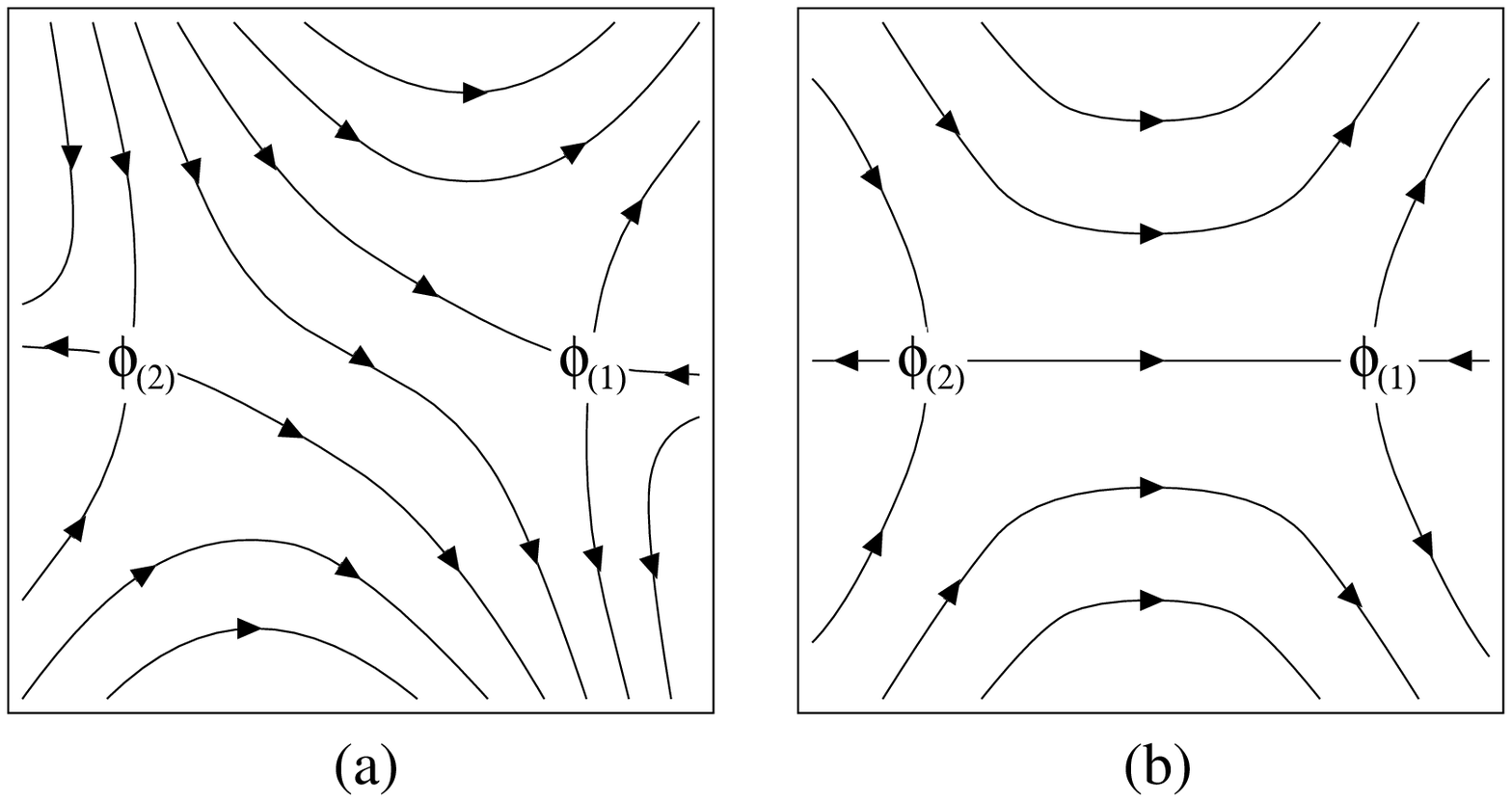}
  \caption{Gradient curves of $H$ for $a_1=-3$. (a) $\tan\beta=1$,
           (b) $\tan\beta=0$.}
  \label{fig_Hgrad_2}
\efig

When $\beta\neq 0$ and $\beta\neq\pi$, $\phi_{(1)}$ is not connected
 with $\phi_{(2)}$ by any gradient curves [Fig.\ref{fig_Hgrad_2}(a)].
All gradient curves that start from or come into stationary points
 extend to infinity.
Therefore Eqs.(\ref{solitoneq_2}) do not have soliton solution.

When $\beta=0$, $H$ becomes [Fig.\ref{fig_Hgrad_2}(b)]
\be
 H=-\f 13{\phi_1}^3-a_1\phi_1{\phi_2}^2+\phi_1 .  \label{H3tan0H_2}
\ee
One of the gradient curve connects $\phi_{(1)}$ and $\phi_{(2)}$.
Thus there is a soliton solution.
When $\beta=\pi$, the result is the same as this.

However when $a_1=-1$, the situation is different.
In this case $H(\phi_1,\phi_2)$ becomes harmonic function,
 and it has a conjugate function $h(\phi_1,\phi_2)$
 defined by Cauchy-Riemann equations.
Therefore we can define a new phase $\alpha$ between these functions.
Subsequently N=1 supersymmetry is promoted to N=2.
If $H(\phi_1,\phi_2)$ has the same form as $W_R$ (\ref{W_R})
 when $(\phi_R,\phi_I)$ is replaced by $(\phi_1,\phi_2)$,
 its conjugate function $h(\phi_1,\phi_2)$ is given by $W_I$ (\ref{W_I}).

Then we can rewrite the scalar potential in (\ref{action_2})
\be
 \sum_{i=1}^2\f 12(H_i)^2=\sum_{i=1}^2\f 12(H_i\cos\alpha+h_i\sin\alpha)^2
 \label{deformV_2-2}
\ee
and (\ref{deform_2}) become
\be
&&{\cal H}=\int dx[\ \sum_{i=1}^2\f 12\{\p_x\phi_i
   -(H_i\cos\alpha+h_i\sin\alpha)\}^2  \nn \\ 
&& \hs{3cm} +\p_x(H\cos\alpha+h\sin\alpha)\ ] .
\ee
The mass bound is
\be
 {\cal H}\geq T_{(i,j)}\cos\alpha+t_{(i,j)}\sin\alpha ,
\ee
 where $t_{(i,j)}$ is
\be
 t_{(i,j)}=h(\phi_{(j)})-h(\phi_{(i)}) .
\ee
This bound is saturated by solutions of equations
\be
 \p_x\phi_1=\f\p{\p\phi_1}(H\cos\alpha+h\sin\alpha)   \\
 \p_x\phi_2=\f\p{\p\phi_2}(H\cos\alpha+h\sin\alpha) \label{solitoneq_hal}
\ee
When $\alpha=\beta$, $H\cos\alpha+h\sin\alpha$ is the same
 as $H$ with $\beta=0$.
When $\alpha=\beta+\pi$, it is the same as $H$ with $\beta=\pi$.
That is, by adjusting $\alpha$, we can always put $\beta=0$ or $\beta=\pi$.
Thus when $a_1=-1$, equations of motion have always soliton solutions.
These discussions are consistent with Sec. \ref{N2sol}.
This means that N=1 supersymmetry is promoted to N=2 when $a_1=-1$.

We summarize ranges where equations of motion have soliton solution.
They are $a_1>0$, $a_1=-1$, and $a_1<0$ with $a_2=0$.

Let us notice that
 we can define the conjugate function of $H$ for arbitrary $a_1$,
 although they do not satisfy Cauchy-Riemann equations.
It is derived from $H$ (\ref{H3tanH_2})
 by replacing $\beta$ to $\beta-\f\pi 2$.
\begin{eqnarray}
&&h=-\f 13\sin\beta{\phi_1}^3-\cos\beta{\phi_1}^2\phi_2
-a_1\sin\beta\phi_1{\phi_2}^2-\f 13a_1\cos\beta{\phi_2}^3 \nn  \\
&&\hs{2cm} +\sin\beta\phi_1+\cos\beta\phi_2
\end{eqnarray}
We can introduce the phase $\alpha$ between (\ref{H3tanH_2}) and this,
 and $H\cos\alpha+h\sin\alpha$ is same as $H$ with $\beta=0$ (\ref{H3tan0H_2}).
This construction is similar to the harmonic case.
But since they do not satisfy Cauchy-Riemann equations (\ref{Cauchy-Riemann}),
 we can not deform the scalar potential like (\ref{deformV_2-2}),
 and we can not put $\beta=0$ or $\beta=\pi$ by $\alpha$.
\section{Summary}
     
We considered the soliton in two dimensional N=1 supersymmetric theory
 in which two fields couple through quasi-superpotential.
In this case, masses of soliton solutions
 satisfy combination rule
\begin{eqnarray*}
 M_{12}+M_{23}=M_{13} .
\end{eqnarray*}

We demonstrated this result along with special two parameter family
 of superpotential (\ref{H3_2}).
However, this combination rule of soliton masses is considered
 a general feature of non-harmonic superpotential.

This implies marginal stability of solitons,
 and suggests the absence of intersoliton force
 between some solitons in this model.
To check it by several ways will be our future work.
\section*{Acknowledgements}
     
The author would like to thank M.Yamada for stimulating conversation.
Useful discussions with A.Yokota and N.Motoyui are acknowledged.

     \addcontentsline{toc}{section}{\protect\numberline{ }{Acknowledgements}}

\end{document}